\documentclass[twocolumn,showpacs,preprintnumbers,amsmath,amssymb]{revtex4}
\usepackage{amssymb}
\usepackage{amsmath}
\usepackage{txfonts}

\usepackage{graphicx}
\usepackage{dcolumn}
\usepackage{bm}

\begin{document}


\title{Bound-state QED Solutions of the photons' off-shell propagating behavior in atoms}

\author{Wen-Zhuo Zhang and Wu-Ming Liu}
\affiliation{Beijing National Laboratory for Condensed Matter Physics, Institute of Physics, Chinese Academy of Sciences, Beijing 100190, P.R. China}

\begin{abstract}
We use the S-matrix formalism of bound-state QED to study the photon-atom scattering. We find that the internal lines in Feynman diagrams which describing the propagation of off-shell bound electrons provide the off-shell amplitudes of photons' propagation in atoms phenomenally. Our work set up the connection between the property of Feynman propagators in bound-state QED and the superluminal but casual propagating behavior of light in atomic media. We also studied the relation between the bound-state QED and the widely used light-atom interacting model in quantum optics, and give the experimental condition where only the bound-state QED is valid.
\end{abstract}


\maketitle

\section{Introduction}

The propagating behavior of photons in optical media is interesting due to a refracted photon has same energy but different momentum to it has in vacuum. In quantum field theory \cite{QFT-book}, the two words \emph{on-shell} and \emph{off-shell} differentiate whether a particle's momentum and energy obeying or disobeying the relativistic energy-momentum relation (mass shell). For an on-shell particle, its momentum and energy obeys the relativistic relation $H_0=\sqrt{P^2c^2+m^2c^4}$. For an off-shell particle, its momentum and energy disobey the relativistic relation, which means $H_0\neq\sqrt{P^2c^2+m^2c^4}$. When a photon is propagating in vacuum, it is on-shell due to it obeys the relation $h\omega_0=\hbar k_0c$. When a photon is propagating in an optical medium, its frequency is same to it has in vacuum, however, its wave vector is changed by the refraction. The Abraham-Minkowski controversy \cite{AM1,AM2,AM3,AM4,AM5} is a century-old problem which debates that the momentum of a photon in the optical media is $\hbar k_0/n$ (Abraham momentum) or $n\hbar k_0$ (Minkowski momentum), where $n$ is the refraction index of the optical media. No matter which momentum is right, the refracted photon seems to be always off-shell in optical media due to $h\omega_0/c\neq n\hbar k_0$ and  $h\omega_0/c\neq\hbar k_0/n$

This article is intend to solve the off-shell propagating behavior of photons in atoms during the photon-atom scattering processes. The widely used light-atom interacting models in quantum optics \cite{QO1,QO2} are the semi-classical models and the quantum models such as Jaynes-Cummings model \cite{JC}. Some scattering processes between laser and atoms (stimulated raman scattering \cite{SRS1,SRS2}, and recoil-induced resonances \cite{RIR}, for examples) have been studied well by them. However, the interaction Hamiltonian of these models $H_I=-e\mathbf{D\cdot E}$ are restricted by the electric dipole approximation (EDA), which requires that the radius of an atom $r$ is much less than the wavelength of the light $\lambda$. Thus these quantum optics models are only valid for the interaction between atoms and long-wavelength light. They can not deal with the case where the wavelength of the light is close to or shorter than the scale of the atoms due to the condition $r\ll\lambda$ is not satisfied. Besides, the electric dipole approximation also restricts the electric dipole and the interaction on a spatial point, which makes the quantum optics models can not give the propagating behavior of photons from one point to another in atoms. Other non-relativistic light-atom(ion) scattering models are also well developed to study the Rayleigh scattering \cite{NRS} and Compton scattering \cite{NCS} between X-rays and bound electrons in atoms. These models have no electric dipole approximation. However, they are mainly based on the partial wave expansion which have no propagators to describe the propagating behavior of photons in atoms during the scattering either.

Quantum electrodynamics (QED) is the relativistic quantum field theory for electromagnetic interaction \cite{QFT-book,QED}. It has very accurate predictions and are accepted as the fundamental theory of all light-matter interacting phenomena \cite{QED_S}. The bound-state QED is the extension of QED. It studies the interaction between bound-electrons and photons. The early bound-state QED models are developed by Furry \cite{Furry} to study the bound-state wave function in positron theory, and by Salpeter and Bethe \cite{BS,GL} to study the bound-states of two Fermi-Dirac particles with the S-matrix formalism. In the past twenty-five years, many bound-state QED models were developed to study the electron structures of atoms and ions \cite{Buch,correction1,correction2}, including the QED corrections of energy levels of few-electron atoms (or ions) \cite{FE}, and many-electron atoms (or ions) \cite{ME,ME1,ME2}. The bound-state QED is suitable to be the better theory of photon-atom scattering due to (i) it can deal with the interaction between atoms and the light at all wavelength; (ii) the off-shell propagating behavior of photons in atoms can appear naturally in the S-matrix formalism of bound-state QED.

In this paper, we use the bound-state QED to study the photon-atom scattering, and give the origin of photon's off-shell behavior in atoms during the scattering. Our paper is arranged as follows: In Section II, we present the S-matrix formalism of bound-state QED for the photon-atom scattering processes. In Section III, we study the Feynman propagator of bound electrons in photon-atom scattering and present its off-shell amplitudes, which can give the off-shell amplitudes of photons in atoms phenomenally. Section IV shows the relation between the perturbation theory of bound-state QED and the widely used photon-atom interacting model in quantum optics textbooks, and Section V gives the experiment condition where only the bound-state QED approach of photon-atom scattering is valid and can be tested. Section VI is the summary and conclusion.

\section{Bound-state QED for photon-atom scattering}

The Lagrangian density of quantum electrodynamics (QED) is
\begin{equation}
L_{QED}=\bar{\psi}(i\gamma^{\mu}\partial_\mu-m)\psi+ie\bar{\psi}A_{\mu}\psi+\frac{1}{4}F_{\mu\nu}F^{\mu\nu},\label{LQED}
\end{equation}
with $\psi$ being the Dirac field, $\gamma^{\mu}$ being the Dirac matrix, $A_{\mu}$ being the four-potential of the electromagnetic field, $F_{\mu\nu}=\partial_{\mu}A_{\nu}-\partial_{\nu}A_{\mu}$ being the electromagnetic field tensor, and $e$ being the elementary charge. We use the natural units $\hbar=c=1$ throughout this and next section.

The Lagrangian density of the bound-state QED is usually written
\begin{equation}
L_B=\bar{\psi}_D(i\gamma^{\mu}\partial_\mu+\gamma^0U-m)\psi_D+ie\bar{\psi}_DA_{\mu}\psi_D+\frac{1}{4}F_{\mu\nu}F^{\mu\nu},\label{LB}
\end{equation}
where $U$ is the potential created by the nuclei \cite{FE}, and $\psi_D$ is the bound-state Dirac field which describes the bound electrons in atoms.
The perturbation theory of QED studies the scattering between electrons, positrons, and photons with S-matrix. The perturbation theory of bound-state QED, however, studies the scattering between bound-electrons and photons. Since the nuclei and electrons are bound together in atoms, such scattering is equivalent to the scattering between atoms and photons. In this paper, we use the bound-state QED to study the photon-atom scattering. Since no positrons exist in atoms, we simplify the Furry picture \cite{Furry} by removing the creation and annihilation operators of positrons. Then only the creation and annihilation operators of electrons are preserved. The bound-state Dirac field in external potential $U$ can be canonically quantized as
\begin{equation}
\begin{aligned}
\psi_D&=\sum_{p,n}U_p\phi_n(x)e^{ip\cdot x}b_{p,n}\\
\bar{\psi}_D&=\sum_{p,n}U_p\bar{\phi}_n(x)e^{-ip\cdot x}b^+_{p,n},
\end{aligned}
\end{equation}
where $\phi_n(x)$ is the bound-state wave function of electrons in the energy level $n$ (internal states), $p$ is the linear four-momentum of the bound electron, $x=(\mathbf{x},t)$ is the coordinate of Minkowski space, and $U_p$ is the normalization coefficient of bound electron field with four-momentum $p$ in the potential $U$. Here $b^+_{p,n}$/$b_{p,n}$ is the creation/annihilation operator of the bound electrons, which obeys the anti-commutation relation
\begin{equation}
\begin{aligned}
\{b^+_{p,n},b_{p',n'}\}&=\delta_{nn'}\delta_{pp'}\\
\{b^+_{p,n},b^+_{p',n'}\}&=\{b_{p,n},b_{p',n'}\}=0,\label{AC}
\end{aligned}
\end{equation}
with $\delta$ being the Dirac delta function. Fortunately the angular momentum of a bound electron is fixed at a certain energy level in atoms, thus we do not need to sum the possible spin directions of a bound electron.

The standard canonical quantization of free electromagnetic field is \cite{QFT-book,QED}
\begin{equation}
A_{\mu}=\sum_k\sum_r\left(\frac{1}{2V\omega_k}\right)^{1/2}(\epsilon_ra^+_ke^{-ik\cdot x}+\epsilon_ra_ke^{ik\cdot x}),
\end{equation}
with $a^+_k/a_k$ being the creation/annihilation operator of the photons with four-momentum $k$, which obeys the commutation relation
\begin{equation}
\begin{aligned}
{[}a^+_k,a_{k'}]&=\delta_{kk'}\\
[a^+_k,a^+_{k'}]&=[a_k,a_{k'}]=0.
\end{aligned}
\end{equation}
Here $\epsilon_r$ is the polarization vector of the photons, and $V$ is the quantization volume whose surface has periodic boundary conditions for the electromagnetic field \cite{QFT-book,QED}.

Then the S-matrix formalism
\begin{equation}
\begin{aligned}
S&=\sum_n^\infty S^{(n)}\\
&=\sum_n^\infty\frac{(-i)^n}{n!}\int...\int d^4x_1... d^4x_n[H_I(x_1)...H_I(x_n)]\label{S}
\end{aligned}
\end{equation}
with interaction Hamiltonian density
\begin{equation}
H_I(x)=e\textbf{N}\{\bar{\psi}_D(x)A_{\mu}(x)\psi_D(x)\}
\end{equation}
can be applied to the study of photon-atom scattering process. Here the operator $\textbf{N}\{\bar{\psi}_D(x)A_{\mu}(x)\psi_D(x)\}$ means the normal product of the $\bar{\psi}_D(x)A_{\mu}(x)\psi_D(x)$.

A feature of the S-matrix is that the initial and final states of it are separated with a large space-time scale where the interaction Hamiltonian can be adiabatically removed \cite{QFT-book}. Thus the initial and final states of S-matrix need to be the unperturbable states of particles. This approach causes a problem to all excited states of bound electrons in atoms since the perturbation from the vacuum state of the electromagnetic field can make all the excited states of bound electrons decay in finite time. Such phenomenon is the spontaneous emission which is based on the Weisskopf-Wigner theory \cite{WW}. In previous bound-state QED approaches that focus on the QED correction of the energy levels of atoms, the coupling between the excited states of bound electrons and the vacuum state of electromagnetic field are rationally neglected \cite{Buch}, then all internal states of atoms are considered as stationary unperturbed states.

However, for the photon-atom scattering, the coupling between the bound electrons and the vacuum state of the electromagnetic field cannot be neglected. This coupling can always make the excited states of bound electrons being perturbed and decay to the ground states with spontaneous emission during the photon-atom scattering. Thus only the unperturbable ground states of bound electrons are suitable to be the initial and final bound electron states of the S-matrix. The spontaneous emission is just included in the scattering process. For this reason, all the excited states of bound electrons in the photon-atom scattering are treat as off-shell states in this paper. The detail relation between the Weisskopf-Wigner theory and the bound-state QED can be found in Section IV.

With the discussion above, the quantization of Dirac field in external potential $U$ can be simplified as
\begin{equation}
\begin{aligned}
\psi_D&=\sum_pU_p\phi_g(x) e^{ip\cdot x}b_p\\
\bar{\psi}_D&=\sum_pU_p\bar{\phi}_g(x) e^{-ip\cdot x}b^+_p.
\end{aligned}
\end{equation}
Here $\phi_g(x)$ is the bound-state wave function of the electrons in the ground states of atoms. We are interested in the photon-atom scattering processes without any change of the photon number, which can be described by the second order S-matrix.

\begin{figure}
\includegraphics[width=70mm]{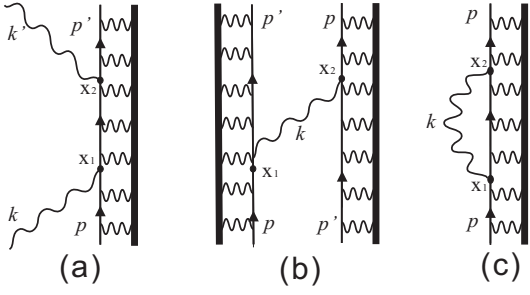}
\caption{Feynman diagrams of the second order S-matrix of light-atom interaction. Thick solid lines denote to the nuclei, thin solid lines denote to the bound electrons, and wave lines denote to the photons. $p$ and $p'$ are the linear four-momentum of the bound electron; $k$ and $k'$ are the four-momentum of the photons.}\label{f2}
\end{figure}

The second order S-matrix formalism can be expanded by the Wick theorem \cite{QFT-book,QED}
\begin{equation}
\begin{aligned}
S^{(2)}&=-\frac{e^2}{2}\int d^4x_1d^4x_2N[\bar{\psi}_D(x_2)A_{\mu}(x_2)\psi_D(x_2)]\\
&\times N[\bar{\psi}_D(x_1)A_{\mu}(x_1)\psi_D(x_1)]\\
&=-\frac{e^2}{2}\int d^4x_1d^4x_2\\
&\times N[\bar{\psi}_D(x_2)A_{\mu}(x_2)\psi_D(x_2)\bar{\psi}_D(x_1)A_{\mu}(x_1)\psi_D(x_1)\\
&+\bar{\psi}_D(x_2)A_{\mu}(x_2)\underline{\psi_D(x_2)\bar{\psi}_D(x_1)}A_{\mu}(x_1)\psi_D(x_1)\\
&+\bar{\psi}_D(x_2)\psi_D(x_2)\underline{A_{\mu}(x_2)A_{\mu}(x_1)}\bar{\psi}_D(x_1)\psi_D(x_1)\\
&+\bar{\psi}_D(x_2)\underline{A_{\mu}(x_2)\psi_D(x_2)\bar{\psi}_D(x_1)A_{\mu}(x_1)}\psi_D(x_1)\\
&+\underline{\bar{\psi}_D(x_2)A_{\mu}(x_2)\psi_D(x_2)\bar{\psi}_D(x_1)A_{\mu}(x_1)\psi_D(x_1)}].\label{S2}
\end{aligned}
\end{equation}
Fig.~\ref{f2} shows the Feynman diagrams corresponding to three processes that described by Eq.~(\ref{S2}). Fig.~\ref{f2}(a) is the Feynman diagram of light-bound electron scattering, which is equivalent to the photon-atom scattering due to the interaction between nuclei and electrons can make the linear four-momentum of the bound electrons become the four-momentum of the atoms' central-of-mass motion eventually. Fig.~\ref{f2}(b) is the Feynman diagram of two bound electrons exchanging a virtual photon. Fig.~\ref{f2}(c) is the Feynman diagram of electron's self-interaction.

We focus on Fig.~\ref{f2}(a) for the photon-atom scattering. We assume that the bound electron has unique internal ground state with wave function $\phi_g(x)$. Let the linear four-momentum of the bound electron being $p=\textbf{p}+E_p$, and the four-momentum of the photon being $k=\textbf{k}+\omega_k$. There are three photon-atom scattering processes that can be studied by the Fig.~\ref{f2}(a).

1) In the case of $E_p=E_{p'}$, $\omega_k=\omega_k'$, $\textbf{p}\neq\textbf{p}'$, and $\textbf{k}\neq\textbf{k}'$, the photon-atom scattering exchanges momentum but no energy. This process is corresponding to the Rayleigh scattering, where the propagating direction of a photon is changed by the atom but the energy of the photon is unchanged.

2) In the case of $E_p=E_{p'}$, $\omega_k=\omega_k'$, $\textbf{p}=\textbf{p}'$, and $\textbf{k}=\textbf{k}'$, the photon-atom scattering exchanges neither momentum nor energy. This process is corresponding to the circle of absorption and stimulated emission of single-frequency photons by atoms.

3) In the case of $E_p\neq E_{p'}$, $\omega_k\neq \omega_k'$, $\textbf{p}\neq\textbf{p}'$, and $\textbf{k}\neq\textbf{k}'$, the photon-atom scattering exchanges both momentum and energy. This process is corresponding to the Compton scattering of high-energy photons by atoms. For low-energy photons, this process is also corresponding to the circle of absorption and spontaneous emission of photons by atoms.

In the second case above, the intermediate state (from $x_1$ to $x_2$) can be seen as a photon refracted by an atom due to the photon has the same momentum and energy before $x_1$ and after $x_2$ (corresponding to a refracted light has same momentum and energy before going into and after going out of a optical media). From $x_1$ to $x_2$, the four-momentum $k$ of the photon is carried by the bound electron, where its refracted behavior is determined by the Feynman propagator of this bound electron. We will give this Feynman propagator in next section.

We have assumed that the bound electron has unique ground state. If the bound electron has more than one ground states, all the ground states are suitable to be the initial and final states of S-matrix. Therefore the initial and final states of the bound-electrons in Fig.~\ref{f2}(a) can have different internal wave function with different internal energy. The corresponding quantization of Dirac field in external potential $U$ for more than one ground states is
\begin{equation}
\begin{aligned}
\psi_D&=\sum_{p,m}U_p\phi_m(x)e^{ip\cdot x}b_{p,m}\\
\bar{\psi}_D&=\sum_{p,m}U_p\bar{\phi}_m(x)e^{-ip\cdot x}b^+_{p,m},
\end{aligned}
\end{equation}
with $m$ being the number of the ground states. In such cases, the \emph{Raman scattering} of light by atoms can be studied by Fig.~\ref{f2}(a), with the initial and final bound electrons being on different $\phi_m(x)$.

\section{Feynman propagators of bound electrons and the propagating behavior of photons in atoms}

\begin{figure}
\includegraphics[width=50mm]{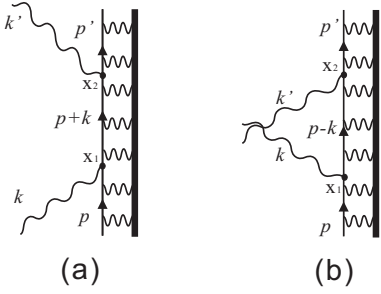}
\caption{Feynman diagrams of the S-channel (a) and U-channel (b) of the photon-atom scattering. Thick solid lines denote to the nuclei, thin solid lines denote to the bound electrons, and wave lines denote to the photons. $p$($p'$) is the linear four-momentum of the bound electron and $k$($k'$) is the four-momentum of the photon.}\label{f3}
\end{figure}

There are two channels of Fig.~\ref{f2}(a) that describing the photon-atom scattering processes. One is the S-channel (Fig.~\ref{f3}(a)), and the other is the U-channel (Fig.~\ref{f3}(b)). In the S-channel, the bound electron absorbs a photon at $x_1$ and emits a photon at $x_2$ with $x_2>x_1$. In the U-channel, the bound electron emits a photon at $x_1$ and absorbs a photon at $x_2$ with $x_2>x_1$.

With the initial state of Fig.~\ref{f3}(a) being $|i>=b^+_pa^+_k|0>$, and the final state of Fig.~\ref{f3}(a) being $|f>=b^+_{p'}a^+_{k'}|0>$, the corresponding S-channel S-matrix element can be written
\begin{equation}
\begin{aligned}
<f|S^{(2)}_a|i>&=-e^2\int d^4x_1x_2\bar{\phi}_g(x_2)U_{p'}e^{ip'x_2}\left(\frac{1}{2V\omega_{k'}}\right)^{1/2}e^{ik'x_2}\\
&\times\frac{1}{(2\pi)^4}\int d^4(p+k)e^{-i(p+k)\cdot(x_2-x_1)}iS_F(p+k)\\
&\times\phi_g(x_1)U_pe^{ipx_1}\left(\frac{1}{2V\omega_k}\right)^{1/2}e^{ikx_1}.\label{SC}
\end{aligned}
\end{equation}
Here $S_F(p+k)$ is the Feynman propagator of the bound electron with linear four-momentum $(p+k)$. Because the angular momentum of a bound electron is fixed at a certain energy level, only the photons with certain polarizations can be absorbed and emitted by the bound electron in the transition between two energy levels. Therefore we do not need to sum the polarization of the photons. The integrating result of Eq.~(\ref{SC}) is
\begin{equation}
\begin{aligned}
<f|S^{(2)}_a|i>&=(2\pi)^4\delta^{(4)}(p'+k'-p-k)\\
&\times\sqrt{\frac{1}{4V^2\omega_{k'}\omega_k}}U_{p'}U_p\cdot[-e^2iS_F(p+k)]\\
&\times\int d^4x_2\bar{\phi}_g(x_2)\int d^4x_1\phi_g(x_1).
\end{aligned}
\end{equation}
Similarly, the U-channel S-matrix element is
\begin{equation}
\begin{aligned}
<f|S^{(2)}_b|i>&=(2\pi)^4\delta^{(4)}(p'-k'-p+k)\\
&\times\sqrt{\frac{1}{4V^2\omega_{k'}\omega_k}}U_{p'}U_p\cdot[-e^2iS_F(p-k)]\\
&\times\int d^4x_2\bar{\phi}_g(x_2)\int d^4x_1\phi_g(x_1),
\end{aligned}
\end{equation}
with $S_F(p-k)$ being the Feynman propagator of the bound electron with linear four-momentum $(p-k)$.

For the S-channel, the photon is absorbed at space-time point $x_1$, and emitted at $x_2$. The property of the photon's propagation in the atom is determined by the Feynman propagator of the bound electron between $x_1$ and $x_2$, which is $S_F(p+k)$. For the reason that the initial and final states of the S-matrix need to be on-shell states, the on-shell state of the bound electron is its the ground state with internal wave function $\phi_g$ and arbitrary linear four-momentum $p=\textbf{p}+\sqrt{\textbf{p}^2+m^2}$, while the state of the bound electron between $x_1$ and $x_2$ is its off-shell state due to $\sqrt{(\textbf{p}+\textbf{k})^2+m^2}\neq\sqrt{\textbf{p}^2+m^2}+\omega_k$. Therefore, the $S_F(p+k)$ gives the off-shell amplitude of the bound electron field propagating from $x_1$ to $x_2$, which can be phenomenally considered as the photon propagating in the atom from $x_1$ to $x_2$ with such off-shell amplitude.

There are interesting results that caused by the Feynman propagator of the bound electron field. In momentum-space, the propagator is
\begin{equation}
S_F(p+k)=\frac{1}{\gamma^{\mu}(p+k)_\mu+\gamma^0U-m+i\epsilon}.
\end{equation}
In position space, the propagator has the form
\begin{equation}
S_F(x_2-x_1)=\int\frac{d^4(p+k)}{2\pi^4}\cdot\frac{e^{-i(p+k)(x_2-x_1)}}{\gamma^{\mu}(p+k)_\mu+\gamma^0U-m+i\epsilon},\label{FP}
\end{equation}
which is different to the Feynman propagator of a free electron field by a potential $\gamma^0U$ appearing in the denominator.

We know that in QED, the Feynman propagator of the free electron field is nonzero outside the light cone although it falls rapidly for space-like intervals \cite{QFT-book,QED}. It means that the off-shell free electron has the superluminal amplitude. However, all operators commute with each other at space-like separation in quantum field theory, which makes such superluminal amplitude never violate the causality \cite{QFT-book,QED}.

This case is also suitable to the Feynman propagator of bound electrons. In Eq.~(\ref{FP}), the integrating result is nonzero when $x_1$ and $x_2$ are space-like separated. This result means that the bound electron field has the superluminal amplitude just as the free Dirac field has. Another way to see it more clearly is that the potential $\gamma^0U$ is independent to $(p+k)$, then it can be combined with $m$ into $(m-\gamma^0U)$ in the integration. Thus the Feynman propagator of the bound electron in Eq.~(\ref{FP}) is very similar to the free electron's one $S_F(x_2-x_1)=\int\frac{d^4(p+k)}{2\pi^4}\cdot\frac{e^{-i(p+k)(x_2-x_1)}}{\gamma^{\mu}(p+k)_\mu-m+i\epsilon}$, only with $m$ being replaced by $(m-\gamma^0U)$. The two propagators have similar superluminal amplitude which never violate the causality.

\begin{figure}
\includegraphics[width=70mm]{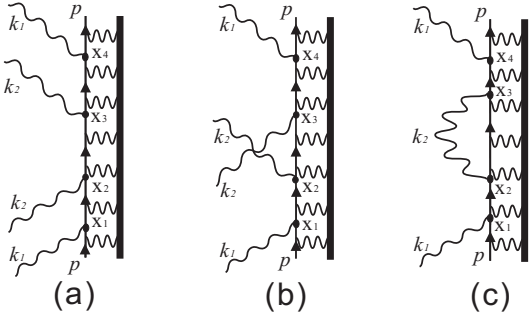}
\caption{Fourth order Feynman diagrams of photon-atom scattering in pump-probe configuration. Thick solid lines denote to the nuclei, thin solid lines denote to the bound electrons, and wave lines denote to the photons. $p$ is the linear four-momentum of the bound electron, and $k_1$ is the four-momentum of the photon from probe light, and $k_2$ is the four-momentum of the photon from pump light.}\label{f4}
\end{figure}

There are works on observing the superluminal group velocity of light pulse in atomic vapor with pump-probe configurations \cite{ftl1,ftl2}. The probe pulse has superluminal group velocities in the atom vapor within the abnormal dispersion frequency range. The process, that a probe laser pulse going into the pumped atomic vapor and going out of it in a short time, is just a good case of photon-atom scattering. It can be described by higher order S-matrix that corresponding to the Feynman diagrams in Fig.~\ref{f4}, with photons from both pump ($k_2$) and probe light ($k_1$) being scattered by the bound electron. The Feymnan propagator of the bound electron form $x_1$ to $x_4$ in Fig.~\ref{f4} is
\begin{equation}
\begin{aligned}
S_F(x_4-x_1)&=\int\frac{d^4(p+k_1)}{2\pi^4}\cdot\frac{e^{-i(p+k_1)(x_2-x_1)}}{\gamma^{\mu}(p+k_1)_\mu+\gamma^0U-m+i\epsilon}\\
&\times\int\frac{e^{-i(p+k_1\pm k_2)(x_3-x_2)}d^4(p+k_1\pm k_2)}{2\pi^4(\gamma^{\mu}(p+k_1\pm k_2)_\mu+\gamma^0U-m+i\epsilon)}\\
&\times\int\frac{d^4(p+k_1)}{2\pi^4}\cdot\frac{e^{-i(p+k_1)(x_4-x_3)}}{\gamma^{\mu}(p+k_1)_\mu+\gamma^0U-m+i\epsilon}.\label{PP}
\end{aligned}
\end{equation}
We see this Feynman propagator can give the superluminal but causality amplitude as same as Eq.~(\ref{FP}). Phenomenally, it can be seen as  the probe photon obtianing the superluminal but causality amplitude from $x_1$ to $x_4$. Although it was studied by Kramers-Kronig relation that the superluminal group velocity of light pulse in pumped atomic vapor obeys the causality \cite{cau1,cau2}, the bound-state QED here gives a more fundamental reason of obeying the causality in such superluminal phenomenon, which directly comes from the causality of the bound electron's Feynman propagators.

Another interesting result is about the U-channel of photon-atom scattering (Fig.~\ref{f3}(b)). It describes an atom emitting a photon at $x_1$ then absorbing one back at $x_2$. The off-shell bound electron here carries the four-momentum ($p-k$), whose energy is below the energy of the bound electron in ground state. This process recalls the ``rotating wave approximation" in quantum optics textbooks. The relationship between the bound-state QED and the light-atom interacting model of quantum optics will be found in next Section, where more details about the connection between Fig.~\ref{f3}(b) and the rotating wave approximation are presented.

\section{Relations between the perturbation theory of Bound-state QED and the electric dipole approximation}

Our start point is the Schr\"{o}dinger equation of bound-state QED in interaction picture, which can be written
\begin{equation}
\frac{\partial}{\partial t}|\Psi(x)>=-iH_I(x)|\Psi(x)>.\label{IP}
\end{equation}
Here $x=(\mathbf{x},t)$ is the coordinate of Minkowski space, $\Psi(x)$ is the particle number state, and $H_I(x)=ie\bar{\psi}_D(x)A_{\mu}(x)\psi_D(x)$ is the interaction Hamiltonian density of bound-state QED. We use the natural units $\hbar=c=1$ here.

There are two approaches to solve Eq.~(\ref{IP}). One is the perturbation theory of bound-state QED, the other is the widely used light-atom interaction model that based on the electric dipole approximation.

The bound-state QED approach solves Eq.~(\ref{IP}) by the perturbation theory
\begin{equation}
\begin{aligned}
|\Psi(\mathbf{x},t)>&=|\Psi(\mathbf{x},t_0)>-i\int_{t_0}^tdt_1H_I(t_1)|\Psi(\mathbf{x},t_1)>\\
&=|\Psi(\mathbf{x},t_0)>-i\int_{t_0}^tdt_1H_I(t_1)|\Psi(\mathbf{x},t_0)>\\
&+(-i)^2\int_{t_0}^tdt_1\int_{t_0}^{t_1}dt_2H_I(t_1)H_I(t_2)|\Psi(\mathbf{x},t_0)>\\
&+...
\end{aligned}
\end{equation}
Then the S-matrix is defined as
\begin{equation}
S=\lim_{t_0=(-\infty,+\infty)}\frac{|\Psi(\mathbf{x},t)>}{|\Psi(\mathbf{x},t_0)>}.
\end{equation}
All the bound-state QED equations we studied in this paper is based on the S-matrix. We can see the bound-state QED approach cares about the initial and final states of S-matrix, which is an integrating equation of the light-atom interaction.

The widely used light-atom interaction model in quantum optics solves Eq.~(\ref{IP}) by the electric dipole approximation \cite{QO1,QO2}, which is a combination of two approximations. The first is the dipole approximation, which requires the wavelength of light is much larger than the diameter of the atom $\mathbf{k\cdot r}\ll1$. Then the vector potential of the light $\mathbf{A}(\mathbf{x+r},t)$ can be written
\begin{equation}
\begin{aligned}
\mathbf{A}(\mathbf{x+r},t)&=\mathbf{A}(t)\exp(i\mathbf{k\cdot x})(1+i\mathbf{k\cdot r}+...)\\
&\simeq \mathbf{A}(t)\exp(i\mathbf{k\cdot x}).
\end{aligned}
\end{equation}
Usually, $\mathbf{x}$ is set to zero for convenience. The second approximation neglects the coupling between the atoms and the magnetic field of the light due to it is much smaller than the coupling between the atoms and the electric field of the light. Then the electric dipole approximation can be made by defining a new state of the system $\Phi(x)$ as
\begin{equation}
|\Psi(\mathbf{x},t)>=\exp[ie\mathbf{A}(\mathbf{x},t)\cdot D]|\Phi(\mathbf{x},t_0)>,\label{NW}
\end{equation}
where $D$ is the electric dipole formed by the nuclei and the bound electron that absorbing and emitting the light. With Eq.~(\ref{NW}), and $\dot{\mathbf{A}}(\mathbf{x},t)=\mathbf{E}(\mathbf{x},t)$, the Eq.~(\ref{IP}) becomes
\begin{equation}
\frac{\partial}{\partial t}|\Phi(x)>=-e\mathbf{E}(\mathbf{x},t)\cdot D|\Phi(x)>.
\end{equation}
Therefore, the interaction Hamiltonian in electric dipole approximation is
\begin{equation}
H_I=-e\mathbf{E}(\mathbf{x},t)\cdot \mathbf{D},\label{DE}
\end{equation}
which is the primarily used interaction Hamiltonian in quantum optics.

Let us compare the two approaches. The perturbation theory of bound-state QED deals with the interaction between the quantized bound-electron field and the quantized electromagnetic field. The interaction happens at any arbitrary point in Minkowski space, so the S-matrix needs to integrate the interaction Hamiltonian density $H_I(x)=ie\bar{\psi}_D(x)A_{\mu}(x)\psi_D(x)$ over all the points $x$ in Minkowski space \cite{QFT-book,QED}. Such approach requires that the initial and final states are the eigenstates of each field and are separated with infinite distances in space-time, which are just the scattering that can be described by the S-matrix formalism. Off-shell electrons appear naturally in the S-matrix formalism, which are corresponding to the inner line of Feynman diagrams. Due to the integration over the space-time coordinate, the perturbation theory of bound-state QED is suitable to deal with the photon-atom scattering rather than to handle the interaction at a certain point in space-time coordinate.

The electric dipole approximation approach enables bound-state QED to deal with the interaction at a certain point in space-time coordinate. For $\mathbf{k\cdot r}\ll1$, the scale of the atom ``$\mathbf{r}$" is approximated to be infinite small, and the interaction happens at the point ($\mathbf{x}=0$) where the atom exactly locates. The wave function of all bound electrons in the atom is approximately focused on such point, with $\sum_n<\phi_n(\mathbf{x}=0)|\phi_n(\mathbf{x}=0)>=1$. Therefore with $H_I(x)=ie\bar{\psi}_D(x)A_{\mu}(x)\psi_D(x)$ being replaced by Eq.~(\ref{DE}) and $|\Psi(x)>$ being replaced by $\Phi(x)$, the Schr\"{o}dinger equation of light-atom interacting under electric dipole approximation becomes
\begin{equation}
\frac{\partial}{\partial t}|\Phi(x)>=ie\mathbf{E}(\mathbf{x},t)\cdot\mathbf{D}|\Phi(x)>,\label{EDA}
\end{equation}
which has non-perturbation solutions. The electric vector of the light field can be quantized at ($\mathbf{x}=0$) as
\begin{equation}
\mathbf{E}(t)=\mathbf{E}_0(a^+e^{-i\omega_L t}+ae^{i\omega_L t}),
\end{equation}
or even can be its classical form in the semi-classical approach \cite{QO1,QO2}. The widely used light-atom interacting models such as Jaynes-Cummings model \cite{JC} are obtained from electric dipole approximation.

The Weisskopf-Wigner theory \cite{WW} is also obtained from electric dipole approximation. It presents that the reason of the spontaneous emission is the interaction between the excited state of the bound electron ($|e>$) and the vacuum state of electromagnetic field $\sum_k|0_k>$. In Weisskopf-Wigner theory, the $t\rightarrow\infty$ limit of the bound electron's state is its ground state $|g>$. Comparing with the perturbation theory of bound-state QED, such $|g>$ can be the unperturbed final state of S-matrix, while $|e>$ is the perturbed off-shell state. With electric dipole approximation, the perturbation of $|e>$ by $\sum_k|0_k>$ happens at a certain point in space-time coordinate. Without electric dipole approximation, the perturbation of $|e>$ by $\sum_k|0_k>$ needs to be integrated over all the points in space-time coordinate, and finally becomes the $t\rightarrow\pm\infty$ limit of the bound electron's state, which is just $|g>$.

\begin{figure}
\includegraphics[width=25mm]{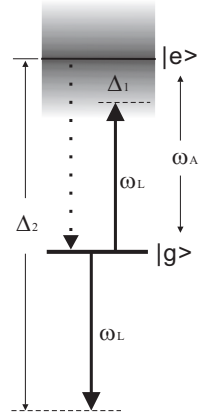}
\caption{Transitions in two-level system. $\omega_L$ is the frequency of light, and $\omega_A$ is the frequency difference between the ground state $|g>$ and the excited state $|e>$. Due to spontaneous emission (dotted line), the excited state has a nature width $\Gamma$. The two transitions $\Delta_1=\omega_A-\omega_L$ and $\Delta_2=\omega_A+\omega_L$ can happen. With rotating wave approximation, the transition $\Delta_2=\omega_A+\omega_L$ are neglected.}\label{level}
\end{figure}

We have mentioned in the end of Section III that Fig.~\ref{f3}(b) is related to the rotating-wave approximation when solving Eq.~(\ref{EDA}). Fig.~\ref{level} is the transition diagram of a two-level atom system interacting with light. Before rotating-wave approximation, both the transitions with detuning $\Delta_1=\omega_L-\omega_A$, and detuning $\Delta_2=-(\omega_L+\omega_A)$ are considered. In the perturbation theory of bound-state QED, the transition loop of stimulated excitation and emission (or spontaneously emission) with detuning $\Delta_1=\omega_L-\omega_A$ is corresponding to the S-channel Feynman diagram of the photon-atom scattering that showed in Fig.~\ref{f3}(a). The transition loop of stimulated excitation and emission with detuning $\Delta_2=-(\omega_L+\omega_A)$ is corresponding to the U-channel Feynman diagram of the photon-atom scattering that showed in Fig.~\ref{f3}(b). After rotating-wave approximation, the transition loop with detuning $\Delta_2=-(\omega_L+\omega_A)$ is neglected. In the perturbation theory of bound-state QED, this is just corresponding to neglecting the contribution from the U-channel Feynman diagram. Since the transition probability of U-channel is much smaller than it of S-channel in S-matrix, the perturbation theory of bound-state QED gives a credible reason for the rotating-wave approximation.

In 2000, Lindgren \emph{et al.} have developed a bound-state QED model based upon a covariant form of the time-evolution operator \cite{TD}. Their work enables the bound-state QED to deal with the time-differential processes by defining the reduced evolution operator $U_{cov}(t,-\infty)$ instead of the S-matrix formalism $S=<f|U_{cov}(\infty,-\infty)|i>$. Their method is free from the electric dipole approximation and can deal with the light-atom interactions besides the photon-atom scattering.

\section{Experimental test}

Since the condition of electric dipole approximation is $\mathbf{k\cdot r}\ll1$, this approximation is only valid in dealing with the interaction between atoms and long-wavelength light, such as ultraviolet, visible, infrared, microwave, \emph{etc}. In the experiments where the wavelength of light is short, such as X-rays, or Gamma rays, the condition $\mathbf{k\cdot r}\ll1$ is no longer satisfied. Thus the electric dipole approximation becomes invalid. However, the perturbation theory of bound-state QED is still valid to explain the phenomena of such experiments due to its interaction Hamiltonian has no approximations that depend on the scale of atoms or the wavelength of light. Therefore, the S-matrix formalism of bound-state QED is suitable to study the experimental phenomena of X-ray photons' off-shell propagating behavior in atoms. For example, the synchrotron radiation or the free-electron lasers that works on X-ray wavelengths can provide clean and coherent X-ray sources to interacting with atoms. A setup of laser cooled and trapped atoms interacting with such coherent X-rays can provide a similar laser-atom interaction platform to them in experimental quantum optics. The off-shell propagating behavior (such as superluminal \cite{ftl1,ftl2}) of X-ray photons in atoms should be observed.

\section{Conclusions}

We use the S-matrix formalism of bound-state QED to study the photon-atom scattering and give the off-shell propagating behavior of photons in atoms during the scattering. By considering the spontaneous emission of excited states of bound electrons, we find that only the bound electrons in ground states are suitable to be the initial and final state of S-matrix. Typical light-atoms scattering processes including Rayleigh scattering, Compton scattering, Raman scattering, as well as the cycle of absorbing and emitting photons by atoms are all corresponding to certain Feynman diagrams in the perturbation theory of bound-state QED. The inner bound electron lines in such Feynman diagrams are described by the Feynman propagators, which can give the off-shell amplitudes of photons in atoms phenomenally. Such propagators have superluminal but causal property, which connects the superluminal propagation of a light pulse in atomic media.

In the bound-state QED, the energy levels of the atoms are determined by the potential $U$, which is composed of the Coulomb potential, the finite size potential from nuclei, or other potentials from other electrons in the atom \cite{FE}. The value of the Feynman propagators in Eq.~(\ref{FP}) and Eq.~(\ref{PP}) are also determined by the value of $U$. Although the superluminal property of these Feynman propagators can be found without integrating them in position space, we hope future study with suitable $U$ may give more precise integrated results of the Feynman propagators of bound-electrons. These results will give the precise description of photon's propagating in atoms.

The widely used light-atom interaction model in quantum optics can be considered as the electric dipole approximation of the bound-state QED, and is suitable to deal with the time-differential processes, while the perturbation theory of bound-state QED with S-matrix formalism $S(\infty,-\infty)$ is more suitable to study the time-integrating processes, such as photon-atom scattering. However, the electric dipole approximation restricts the widely used light-atom interaction model in quantum optics from studying the atoms interacting with short-wavelength light. The perturbation theory of bound-state QED is always valid due to it can be applied to the interaction between atoms and light at all wavelengths.

This work is supported by NSFC under grants Nos. 10874235, 10934010, 60978019, the NKBRSFC under grants Nos. 2009CB930701, 2010CB922904, and 2011CB921502, and NSFC-RGC under grants Nos. 1386-N-HKU748/10.

\end{document}